\documentclass[prc,preprintnumbers,nofootinbib,twocolumn,showpacs]{revtex4}
\usepackage{graphicx}
\usepackage{bm}
\usepackage{amssymb,amsmath}
\usepackage{epstopdf}
\usepackage{epsfig,dcolumn}
\usepackage{array}

\def\sq{\displaystyle{\not} q} 
\def\sp{\displaystyle{\not} p} 
\def\sk{\displaystyle{\not} k} 
\def\sK{\displaystyle{\not} K} 
\def\sP{\displaystyle{\not} P} 
\def\sD{\displaystyle{\not} \Delta } 
\def\se{\displaystyle{\not} \,\varepsilon } 
\def\beqn{\begin{eqnarray}} 
\def\eeqn{\end{eqnarray}} 
\def\nn{\nonumber} 

\begin{document}
\title{New approach to low energy Virtual Compton Scattering and generalized 
polarizabilities of the nucleon}
\author{M. Gorchtein}
\affiliation{Indiana University, Bloomington, IN 47408, USA}
\date{\today}
\begin{abstract}
Virtual Compton scattering off the nucleon (VCS) is studied in the regime of 
low 
energy of the outgoing real photon. This regime allows one to directly access 
the generalized polarizabilities of the nucleon in a VCS experiment. In the 
derivation of the low energy theorem for VCS that exists in the literature, 
the low energy limit taken for virtual initial photons does not match on 
that for real photons, when one approaches the initial photon's mass shell. 
While this problem has for a long time been attributed to the non-analyticity 
of the Compton amplitude with respect to the photon virtuality, I demonstrate 
that it is merely due to an ill-defined low energy limit for VCS, on one hand, 
and to a particular way of constructing the VCS amplitude, use in the 
literature, on the other. 
I provide a uniform description of low energy Compton scattering with real and 
virtual photons by defining a Lorentz-covariant operator sub-basis for 
Compton scattering in that regime, that has six independent structures. 
Correspondingly, six new generalized polarizabilities are introduced in the 
Breit frame. These polarizabilities are defined as continuous functions of 
the photon virtuality and at the real photon point match onto the nucleon 
polarizabilities known from real Compton scattering.
\end{abstract}
\maketitle

\section{Introduction}
Scattering of photons has been among the basic tests used 
to study the structure of matter and to understand the nature of 
light. The experiments of Faraday and theoretical studies of Maxwell 
established the notion of electromagnetic waves light was identified with. 
Compton effect, the phenomenon of the wave length shift of light scattered off 
an electron that is impossible for a classical wave, established the picture of 
particle-wave duality, along with Einstein's theory of the photoeffect. 
With the discovery of the four fundamental interactions and the continuing 
quest for the sub-nucleon structure, Compton scattering with real and virtual 
photons reemerged as a clean way to study this structure experimentally and 
theoretically. The merit of photons as a unversal elementary probe 
of nucleon structure becomes especially emphasized within the framework of 
sum rules and dispersion relations, the connection between scattering of 
photons of low frequency, and absorption of photons of high frequencies 
\cite{ggt,baldin,gdh}. These are 
relations between low energy coefficients, polarizabilities that describe the 
response of the nucleon structure to the quasi-static external electromagnetic 
field, and the nucleon's photoabsorption spectrum. Analyticity, along with 
unitarity of the Compton amplitude play the central role in this 
derivation. \\
\indent
Introduction of polarizabilities is based on the low energy theorem (LET) 
\cite{low}. Electric 
polarizability $\alpha$ and magnetic susceptibility $\beta$ quantify the 
linear response of the nucleon to the incoming photon's electric (magnetic) 
field $\vec E(\vec B)$, respectively. 
The induced electric (magnetic) dipole interacts with the outgoing photon's 
field $\vec E'(\vec B')$ leading to an effective interaction Hamiltonian 
$4\pi\alpha\vec E\cdot\vec E'+4\pi\beta\vec B\cdot\vec B'$. 
Since at low photon energy $\omega$ both fields are $\sim\omega$, one can 
state that generally, the contribution of the 
unknown nucleon structure enters Compton observables at order $\omega^2$. 
Once the known, energy-independent classical Thomson term is separated out, 
the polarizabilities are directly measureable. Here one can already guess a 
potential complication: the polarizabilities arise in a Hamiltonian, rather 
than Lagrangean.
This means that the procedure of low energy expansion is in general frame 
dependent, since the operator is not Lorentz-covariant. Fortunately, for real 
Compton scattering the frame dependence arises as corrections in powers of a 
small quantity $\omega/M$ and at the accuracy at which the low energy expansion 
is truncated, is irrelevant. \\
\indent
This is not generally the case when one of the photons 
is virtual, i.e. originating from electron scattering. The finite initial 
photon "mass" 
$Q^2$ ensures that only outcoming photon's energy vanishes. The 
incoming and outgoing photon energy vanishes simultaneously in Breit frame that 
treats the two photons symmetrically, but not in center-of-mass or laboratory 
frame. Then it is clear that expanding the virtual Compton amplitude in one 
frame or another can differ by terms $\sim Q^2/M^2$.\\
\indent
Low energy theorem for virtual Compton 
scattering (VCS) was introduced through low energy expansion of the VCS 
scattering coefficients in c.m. frame \cite{guichon}. 
The VCS amplitude was decomposed into multipoles that correspond to dipole and 
quadrupole transitions in the initial and final $\gamma N$ state. At low 
outgoing photon energy, there are only ten multipoles that vanish linearly, 
and factoring out this energy dependence, the ten generalized polarizabilities 
(GP's) were introduced. The term "generalized" refers to the fact that these 
GP's are not numbers, but functions of the three-momentum of the virtual 
photon $\vec q$ that is kept fixed, thus they generalize the real Compton 
scattering (RCS) polarizabilities that should thus be just the limit $q\to0$ 
of the GP's. \\
\indent
This correspondence is troublesome, for instance since ten GP's should 
be related to six polarizabilities of RCS \cite{ragusa}. 
The procedure of \cite{guichon} is clearly non-covariant. Implementing Lorentz 
invariance, in particular crossing 
that relates VCS reactions with interchanged initial and final nucleons, 
it was possible to eliminate four out of ten GP's \cite{andreas}, 
equalling the number of independent GP's that should be confronted to the 
RCS polarizabilities. However, only four of the six GP's match onto their RCS 
counterparts, as two of them vanish at the real photon point. This mismatch 
is usually attributed to the non-analyticity of the Compton amplitude as 
function of the photon virtuality $Q^2$ \cite{guichon, andreas}. 
Recalling the special role the analyticity has played in the derivation of the 
nucleon sum rules, it is highly desirable to clarify the origin of such 
non-analytical behavior, if it indeed takes place. 
In fact, non-analytical behavior has to be related to a physical singularity, 
otherwise it is not acceptable in a field theory.\\
\indent
A similar problem was observed some time ago when performing low energy 
expansion of the forward doubly virtual Compton scattering 
\cite{marcbarbaradrechsel}. The authors found a striking result that in this 
case, even the lowest term of the expansion does not match to the classical 
Thomson term. This mismatch was again attributed to the non-analyticity of the 
Compton amplitude with respect to $Q^2$. 
In a recent paper \cite{lexvvcs}, it was shown that this mismatch 
is a pure artifact of an ill-defined low energy limit. In 
\cite{marcbarbaradrechsel}, the LEX is performed around a point that does not 
correspond to nucleon near its physical mass shell. 
In \cite{lexvvcs} it is shown that the LEX for 
VVCS cannot be performed at forward direction, and instead requires non-zero 
nucleon recoil for virtual photons. The new formulation of LET allowed to 
define the low energy coefficients in a continuous way, and to derive 
the generalized sum rules of the nucleon. \\
\indent
In this article, I aim at revising the LET for VCS in order to ensure that the 
Compton amplitude is continuous and analytical function of all its variables. 
In a paper by L'vov et al. \cite{scherer}, 
an alternative approach to analyzing low-energy 
VCS process was proposed. It was based on i) operating with the Compton basis 
written in terms of the electromagnetic field strength tensors, thus using 
a Lorentz covariant description to begin with, and ii) evaluating those 
operators in Breit frame and rewriting them in terms of electromagnetic field 
three-vectors. This approach showed that one does not 
in general need to perform a multipole decomposition of the VCS amplitude, but 
rather use the classical analogy and interprete the scattering coefficients 
that multiply the covariant structures as Breit-frame polarizabilities. 
Unfortunately, that work only included the 
spin-independent part. Another important point that is 
missing in \cite{scherer}, is that the low energy limit of Guichon et al. was 
used, even if implicitly. Correspondingly, neither a complete set of GP's was 
introduced, nor brought in correspondence with the VCS observables that is 
ultimately the principal reason for performing LEX. 
The aim of the present work is in improving on both points.
This study, as well, capitalizes to a large extent on the work by Ragusa 
\cite{ragusa} who introduced the full low energy expansion of the RCS 
amplitude. \\
\indent
The article is organized as follows. 
I will start with defining the VCS kinematics in Section \ref{kinematics}
and rewriting the covariant VCS tensor 
in terms of electromagnetic strength tensors and in Lorentz-covariant form in 
Section \ref{tensor}.
I will then propose the new way to define the low energy limit for 
VCS that is explicitly Lorentz invariant, respects crossing symmetry and 
ensures that the path along which this limit is taken always lies inside the 
physical region in Section \ref{LET}. 
I will perform the low energy expansion (LEX) of the VCS amplitude in Section 
\ref{LEX}, 
relating the tensors to operators involving electromagnetic field three-vecotrs 
in Breit frame, and provide thus the interpretation of the scattering 
coefficients as polarizabilities. Finally in Section \ref{sec:obs}, I 
investigate the relations between the new GP's and the VCS observables.
\section{Virtual Compton kinematics}
\label{kinematics}
I consider the virtual Compton scattering process 
$\gamma^*(q)+N(p)\to\gamma(q')+N(p')$. Its kinematics is described in terms of 
Lorentz scalars 
\beqn
s&=&(p+q)^2=(p'+q')^2=(P+K)^2\nn\\
u&=&(p-q')^2=(p'-q)^2=(P-K)^2\nn\\
t&=&(q-q')^2=(p'-p)^2\nn\\
Q^2&=&-q^2\geq0
\eeqn
where the nucleon and photon average momenta were introduced, 
$P=\frac{p+p'}{2}$, $K=\frac{q+q'}{2}$. The sum of these variables is fixed 
by 
\beqn
s+u+t+Q^2&=&2M^2,
\eeqn
with $M$ the nucleon mass. The above relation implies that only three of them 
are independent, and it is useful to introduce the "crossing" variable 
\beqn
\nu&=&\frac{s-u}{4M}=\frac{PK}{M}=\frac{s-M^2+\frac{t+Q^2}{2}}{2M}
\eeqn
\indent
Each value of $Q^2$, $\nu$ and $t$ can be related to the incoming and 
outgoing photon energy $\omega,\omega'$, respectively, 
and scattering angle $\theta$ that are frame dependent. 
For given $Q^2$, 
one has the three-vector of the virtual photon 
$|\vec q|=\sqrt{\omega^2+Q^2}\equiv q$, while $|\vec q'|=\omega'$. 
In the 
c.m. frame defined by $\vec p+\vec q=0$ 
that was used in \cite{guichon}, one has
\beqn
\omega'&=&\frac{s-M^2}{2\sqrt{s}}\nn\\
\omega&=&\frac{s-M^2-Q^2}{2\sqrt{s}}\nn\\
\cos\theta&=&\frac{\omega}{q}+\frac{t+Q^2}{2q\omega'}
\eeqn
\indent
Alternatively, in this work the nucleon Breit frame will be used. This frame is 
defined by $\vec P=0$. Breit frame is of advantage because it treats the 
photons in a symmetric manner,
\beqn
\omega&=&\omega'=\frac{PK}{P^0}=\frac{M\nu}{\sqrt{M^2-t/4}}
=\frac{s-M^2+\frac{t+Q^2}{2}}{2\sqrt{M^2-t/4}}\nn\\
\cos\theta_{\gamma^*\gamma}&=&\frac{\omega}{q}+\frac{t+Q^2}{2q\omega}
\eeqn
\section{Compton amplitude}
\label{tensor}
\begin{figure}[ht]
\vspace{0.5cm}
{\includegraphics[height=10cm]{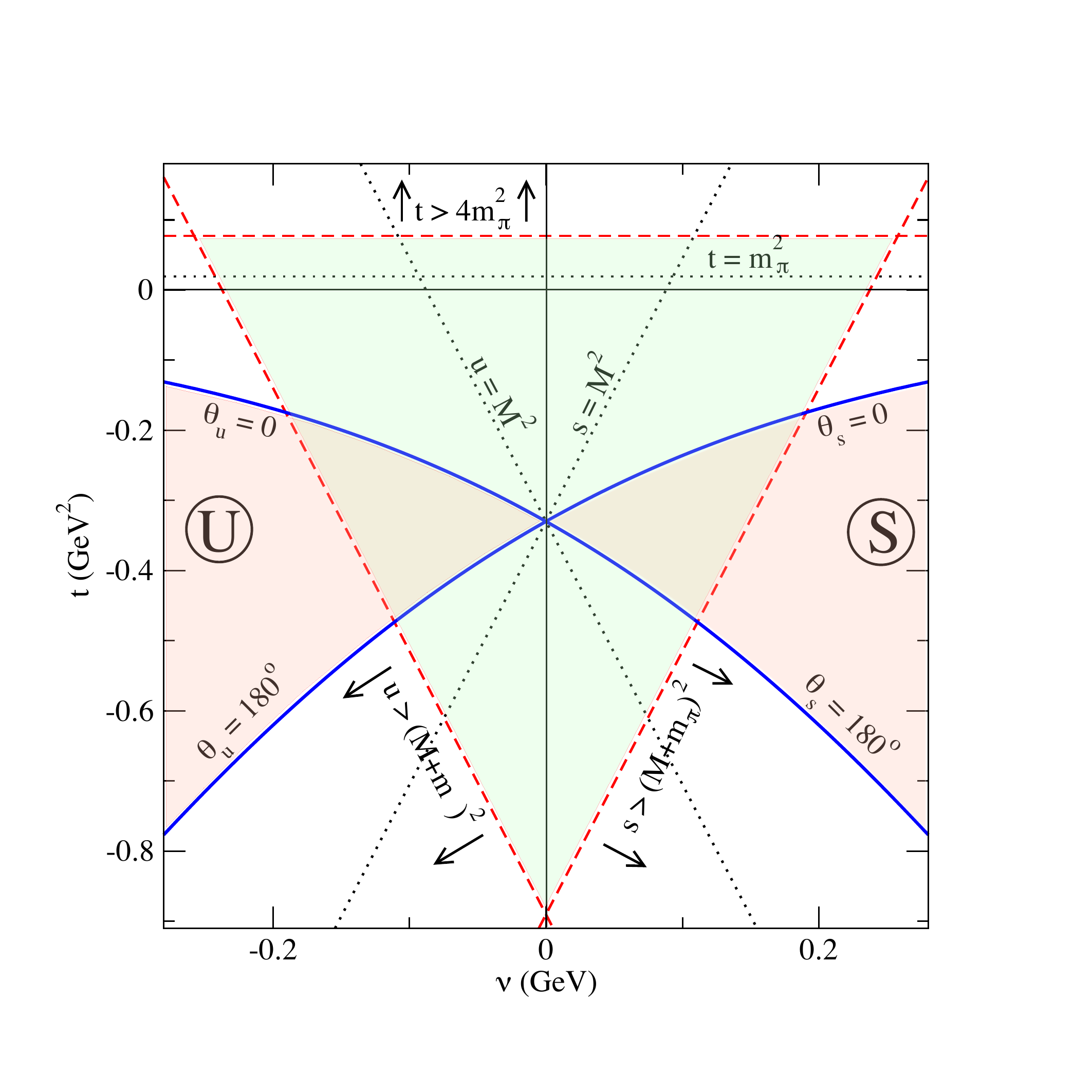}}
\caption{(Color online) Mandelstam plane for VCS with $Q^2=0.33$ GeV$^2$. 
On the plane 
$\nu,t$, the kinematical regions and the positions of the singularities of the 
VCS amplitude are shown. The dotted lines correspond to the nucleon pole in the 
$s(u)$-channel, and $\pi^0$-pole in the $t$-channel. The inelastic thresholds 
are shown by the dashed lines. The scattering regions in the $s(u)$-channels 
correspond to the regions 
between the upper (forward scattering) and lower (backward 
scattering) solid lines at positive (negative) values of $\nu$, respectively.
The VCS amplitude is purely real for all variables below the corresponding 
inelastic threshold (within the triangle). The intersection of the scattering 
regions with the shaded triangle represents the area where the low energy 
expansion can be used.}
\label{fig:mandel}
\end{figure}

\begin{figure}[h]
{\includegraphics[height=3cm]{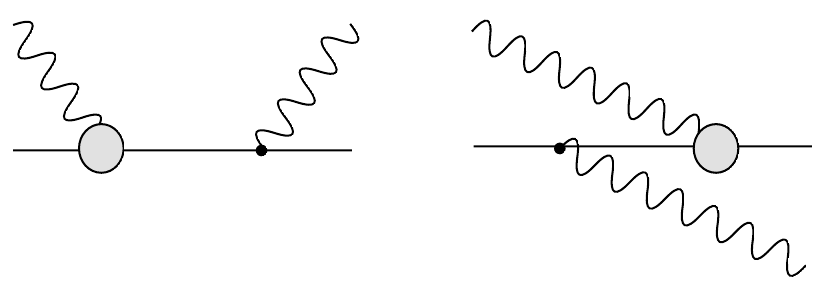}}
\caption{Born contributions to Compton scattering.
Nucleon exchange diagrams in the direct and crosse channels are shown. 
The blobs denote form factors.}
\label{fig:born}
\end{figure}

\begin{figure}[h]
{\includegraphics[height=2cm]{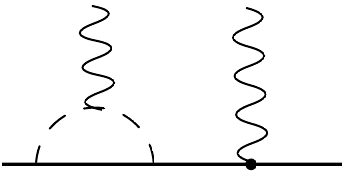}}
\caption{An example of a pion loop contribution to the off-shell nucleon 
form factor. The dashed line denotes the pion loop.}
\label{fig:born1}
\end{figure}

The Lorentz covariant and explicitly gauge invariant tensor basis for VCS 
was introduced long ago \cite{tarrach}. It was then used to study VCS in 
LEX approach \cite{andreas} and in dispersion relations approach \cite{javcs}. 
I will use the form of the VCS tensor of Ref. \cite{tarrach}
\beqn
T^{\mu\nu}_{VCS}&=&\sum_{i=1}^{12}F_i(\nu,t,Q^2)\bar u(p')\rho_i^{\mu\nu}u(p)
\eeqn
as a starting 
point. Above, the $\rho_i$'s are Lorentz covariant tensors that are linearly 
independent and explicitly gauge-invariant by construction \cite{tarrach}. 
The tensors are cast between the initial and final nucleon spinors 
$u(p)$ and $u(p')$, respectively. The numeration of the tensors can be 
different in different references, and for definitiveness I use the one of 
Refs. \cite{andreas}, \cite{javcs}. 
The corresponding amplitudes $F_i$ are Lorentz scalars that are functions of 
the kinematical variables $\nu,t$, and $Q^2$. 
The VCS tensor is then embedded into the full, physical amplitude 
$T_{FVCS}$ for the scattering process $e+p\to e+p+\gamma$ as 
\beqn
T_{FVCS}&=&\frac{e}{Q^2}\bar u(k',h)\gamma_\mu u(k,h)
T^{\mu\nu}_{VCS}\varepsilon'^*_\nu(q',\lambda')\nn\\
&=&\frac{e}{Q^2}\sum_\lambda\Omega(h,\lambda)
\varepsilon_\mu(q,\lambda)T^{\mu\nu}_{VCS}\varepsilon'^*_\nu(q',\lambda'),
\eeqn
where $u(k),u(k')$ denote the initial and final electron's spinors, 
$h$ the conserved helicity of massless electrons, and 
$\Omega(h,\lambda)=\bar u(k',h)\se^*(q,\lambda) u(k,h)$.
I refer the 
reader to the Apendix for the explicit form of the polarization vectors of the 
initial and final photons. 
Above, $e$ stands for the (positron's) electric charge, and my conventions for 
the VCS amplitude differ from those used in the literature 
by a factor of $-e^2$ for further convenience. 
I will next rewrite the twelve tensors in terms of the electromagnetic field 
strength tensors. I define these latter as 
$F^{\alpha\mu}=ie(q^\alpha\varepsilon^\mu - q^\mu\varepsilon^\alpha)$ and 
$F'^{\beta\nu}=-ie(q'^\beta{\varepsilon'^*}^\nu-q'^\nu{\varepsilon'^*}^\beta)$.
The following expressions can be found for 
$\rho_i\equiv\varepsilon_\mu\rho_i^{\mu\nu}\varepsilon'^*_\nu$:
\beqn
\rho_1&=&-\frac{1}{2}F^{\mu\nu}F'_{\mu\nu}\label{eq:tensor1}\\
\rho_2&=&-4\left(P_\mu F^{\mu\alpha}\right)\left(P^\nu F'_{\nu\alpha}\right)
\nn\\
\rho_3&=& \frac{2}{PK}
\left[q^2g_{\alpha\beta}-q_\alpha q_\beta\right]
\left(P_\mu F^{\mu\alpha}\right)\left(P_\nu F'^{\nu\beta}\right)\nn
\eeqn
for the spin-independent part, and 
\beqn
\rho_4&=&2P^\mu
\left[F'_{\mu\alpha}\tilde F^{\alpha\beta}-F_{\mu\alpha}\tilde F'^{\alpha\beta}
\right]i\gamma_5\gamma_\beta\label{eq:tensor2}\\
\rho_5&=&\frac{1}{2}\left[
(q'^\mu F'_{\mu\alpha})\tilde F^{\alpha\beta}
-(q^\mu F_{\mu\alpha})\tilde F'^{\alpha\beta}
\right]i\gamma_5\gamma_\beta\nn\\
\rho_6&=&-\frac{(qq')}{2M}\sD\gamma_5 F^{\mu\nu}\tilde F'_{\mu\nu}\nn\\
&+&2\left[
\left(P_\alpha q_\beta F'^{\alpha\beta}\right)F^{\mu\nu}
-\left(P_\alpha q'_\beta F^{\alpha\beta}\right)F'^{\mu\nu}
\right]i\sigma_{\mu\nu}\nn\\
\rho_7&=&\frac{1}{2}\left[
(q'^\mu F_{\mu\alpha})\tilde F'^{\alpha\beta}
+(q^\mu F'_{\mu\alpha})\tilde F^{\alpha\beta}
\right]i\gamma_5\gamma_\beta\nn\\
\rho_8&=&
\frac{1}{2}\left(q_\mu q_\nu+q'_\mu q'_\nu\right)
F^{\mu\alpha}F'^{\nu\beta}i\sigma_{\alpha\beta}\nn\\
&-&\frac{1}{4}q_\alpha q'_\beta
\left[F^{\alpha\beta}F'^{\mu\nu}+F'^{\alpha\beta}F^{\mu\nu}\right]
i\sigma_{\mu\nu}\nn\\
\rho_9&=&\left[
\left(P_\alpha q_\beta F'^{\alpha\beta}\right)F^{\mu\nu}
+\left(P_\alpha q'_\beta F^{\alpha\beta}\right)F'^{\mu\nu}
\right]i\sigma_{\mu\nu}\nn\\
\rho_{10}&=&-2F^{\mu\alpha}{F'}_\mu^\beta i\sigma_{\alpha\beta}\nn\\
\rho_{11}&=&2\left[
(q^\mu F'_{\mu\alpha})\tilde F^{\alpha\beta}
-(q'^\mu F_{\mu\alpha})\tilde F'^{\alpha\beta}
\right]i\gamma_5\gamma_\beta\nn\\
\rho_{12}&=&-\frac{q^2}{8M}\sD\gamma_5 F^{\mu\nu}\tilde F'_{\mu\nu}\nn\\
&+&\left[
\left(P_\alpha q'_\beta F'^{\alpha\beta}\right)F^{\mu\nu}
-\left(P_\alpha q_\beta F^{\alpha\beta}\right)F'^{\mu\nu}
\right]i\sigma_{\mu\nu}\nn
\eeqn
for the spin-dependent part of the VCS amplitude. 
In the above, the notation $\Delta\equiv q-q'$ was used. 
Transversality condition fixes $q'_\beta F'^{\beta\nu}=0$, while a similar 
condition for the virtual photon is not required. Nevertheless, for the 
reasons of symmetry that will be important in the discussion of the properties 
of the amplitudes $F_i$, I keep terms $\sim q'_\beta F'^{\beta\nu}$ in 
Eq. (\ref{eq:tensor2}). The dual tensor is defined as 
$\tilde F^{\alpha\beta}=\frac{1}{2}\epsilon^{\alpha\beta\mu\nu}F_{\mu\nu}$
and similarly for $F'$. 
One observes that all twelve tensors can be expressed through the field 
strength tensors in a compact way. The first three tensors do not depend on 
the nucleon spin, and coincide with the expressions found in \cite{scherer}. 
The remaining nine tensors are spin-dependent, and were not represented in this 
form in the literature. 
I next consider properties of the VCS amplitudes under two crossing 
transformations: nucleon crossing relates the original reaction 
$\gamma^*(q)+N(p)\to\gamma(q')+N(p')$ to 
$\gamma^*(q)+\bar N(-p')\to\gamma(q')+\bar N(-p)$, while the photon crossing 
- to the process $\gamma(-q')+N(p)\to\gamma^*(-q)+N(p')$. 
Under these transformations, the tensors transform according to
\beqn
P&\to&-P\nn\\
PK&\to&-PK\nn\\
\gamma_5\gamma^\mu&\to&C\gamma_5\gamma^\mu C^\dagger=+\gamma_5\gamma^\mu\nn\\
\sigma^{\mu\nu}&\to&C\sigma^{\mu\nu}C^\dagger=-\sigma^{\mu\nu}
\eeqn
for nucleon crossing and 
\beqn
F^{\alpha\beta}&\leftrightarrow&{F'}^{\alpha\beta}\nn\\
q&\leftrightarrow&-q'\nn\\
K&\to&-K\nn\\
PK&\to&-PK\nn\\
q^2&\leftrightarrow&q'^2
\eeqn
under photon crossing. 
Requiring the VCS amplitude to be invariant under these transformations one 
obtains following properties of the amplitudes \cite{tarrach,andreas}:
\beqn
F_i(-\nu,t,q^2,q'^2)&=&+F_i(\nu,t,q^2,q'^2), \nn\\
&&i=1,2,5,6,7,9,11,12\nn\\
F_i(-\nu,t,q^2,q'^2)&=&-F_i(\nu,t,q^2,q'^2), \nn\\
&&i=3,4,8,10\nn\\
F_i(-\nu,t,q'^2,q^2)&=&+F_i(\nu,t,q^2,q'^2), \nn\\
&&i=1,2,5,6,11,12\nn\\
F_i(-\nu,t,q'^2,q^2)&=&-F_i(\nu,t,q^2,q'^2), \nn\\
&&i=3,4,7,8,9,10
\eeqn
\indent
For both photons real, the tensors $\rho_{3,5,8,12}$ vanish. Furthermore, 
the amplitudes $F_{7,9}$ vanish in that limit. Correspondingly, real Compton 
scattering is described in terms of six amplitudes $F_{1,2,4,6,10,11}$ and 
the corresponding tensors.\\
\indent
Before moving on to discuss low energy behavior of the tensors and the 
amplitudes, the ground state and continuum contributions to the $F_i$'s should 
be separated. On general grounds, the VCS amplitude can have two kinds of 
singularities, poles corresponding to an exchange of an on-shell particle in 
one of the channels, and cuts along which the amplitude has a non-zero 
discontinuity corresponding to multi-particle exchanges. The positions of 
singularities and various kinematical regions for the VCS process are displayed 
in Fig. \ref{fig:mandel}.
The full amplitude can then be splitted in two pieces, out of 
which one would contain poles, and the other one can only have cuts. 
\beqn
T_{VCS,NB}^{\mu\nu}&\equiv&T_{VCS}-T_{VCS,Born}^{\mu\nu}\\
\nn
\eeqn
\indent
The Born (nucleon pole) contribution is due to an exchange of a single 
nucleon in the direct and crossed channel, as shown in Fig. \ref{fig:born}. 
This amplitude is given by
\begin{widetext}
\beqn
T_{VCS,\,Born}^{\mu\nu}&=&
-e^2\bar u(p')\left[
\frac{\Gamma^\nu(q')(\sP+\sK+M)\Gamma^\mu(q)}{(P+K)^2-M^2+i\epsilon}
+\frac{\Gamma^\mu(q)(\sP-\sK+M)\Gamma^\nu(q')}{(P-K)^2-M^2+i\epsilon}
\right]u(p)
\label{eq:born}
\eeqn
\end{widetext}
where the nucleon electromagnetic vertex is given by 
$\Gamma^\mu(q)=F_1(q^2)\gamma^\mu+F_2(q^2)i\sigma^{\mu\alpha}\frac{q_\alpha}{2M}$, and 
$\Gamma^\nu(q')=e_N\gamma^\mu+\kappa_Ni\sigma^{\mu\alpha}\frac{q'_\alpha}{2M}$. 
In the above, $e_N$ and $\kappa_N$ denote the nucleon charge and anomalous 
magnetic moment, whereas 
$F_{1,2}(q^2)$ stand for the usual Dirac and Pauli form factors. For 
practical purposes the form factors are taken in the phenomenological form. 
This choice corresponds to taking out that part of the VCS amplitude that is 
known from other experiments and does not contain any new information.
It should be noticed that this choice is not unique: the phenomenological 
form factors describe the nucleon on its mass shell. Then, in Eq. 
(\ref{eq:born}), it is only the imaginary part that always corresponds to the 
exchange of an on-shell nucleon in either direct or crossed channel. The real 
part of the 
Born amplitude contains off-shell nucleons, and its form factors are in general 
unknown. One may argue that, since the imaginary part in the $s$ and $u$ 
channel is known and given by two $\delta$-functions, a dispersion relation in 
those channels would restore the picture of the real part being given in terms 
of the same on-shell form factors. 
However, such a dispersion representation is incomplete because it neglects 
the analytical structure in the $t$-channel. For instance, Fig. \ref{fig:born1} 
shows a contribution that simultaneously contains the nucleon pole and the 
$\pi N$ continuum. 
The contribution in Fig. \ref{fig:born1} can be obtained from unitarity in 
the $t$-channel for any "mass" of the off-shell nucleon, 
but this cannot be done in a model that uses the 
phenomenological parametrizations of the nucleon form factors. 
Then, this (model-dependent) energy running of the $\gamma^*NN$ vertex 
has to accompany the model-independent part associated with the on-shell 
nucleons. 
The ambiguity arises when the Born and non-Born amplitudes are calculated in 
two different models. Typically, the Born part is evaluated with the 
phenomenological form factors that are used in the analysis of VCS experiments 
(see Section \ref{sec:obs} for details). 
While the divergent $\sim1/\omega$ terms are model-independent, the subleading 
terms $\sim\omega^0,\omega^1$ included in this contribution, are not. 
They can contribute at the same order as the GP's and should be taken with care 
when calculating VCS amplitude in a model, and comparing results to the 
experiment. Any model gives GP's with respect to the Born 
contribution defined and evaluated in the same model.\\
\indent
In the following, I will use the Born amplitude defined in terms of the 
phenomenological form factors, as the most practical choice.\\
\indent
It has been argued, as well, that the $\pi^0$-exchange contribution should 
be included into the Born part since it is a tree-level graph, and it has a 
pole at $t=m_\pi^2$.
There is however an important difference between the nucleon pole graphs and 
the pion pole: the former depend on interaction of a single photon 
with the nucleon, whereas the latter represents a local two-photon interaction, 
and thus contains information complementary to the nucleon pole contributions. 
There is still no consensus in the community, where this contribution should 
be included, so I choose to follow the tradition of attributing the pion 
pole to the continuum contribution. For the purpose of low energy scattering 
in the $s$-channel, this $t$-channel pole lies far enough so that its 
contribution is a continuous function of all variables. \\
\indent
Once the Born amplitude is specified, the Born contributions to the amplitudes 
$F_i$, $F_i^B$ can be calculated. The results are known 
\cite{tarrach,andreas,javcs}, and I will not quote them here. 
The residual part of the amplitude can be generically introduced as 
$F_i^{NB}\equiv F_i-F_i^B$, and the non-Born amplitude contains no poles. 
Therefore, (if one stays 
away from the $t$-channel), this amplitude should be a regular function of all 
its arguments, and the only kind of singularity that it has are the unitarity 
cuts in $\nu$ corresponding to nucleon excitations and continuum in the 
$s$ and $u$ channels. The inelastic thresholds in these channels are separated 
from the 
nucleon pole by the finite pion mass. Then, in the energy range between 
the nucleon pole and the threshold the non-Born amplitude is a purely real 
regular function (see Fig. \ref{fig:mandel}) 
that can be Taylor expanded in powers of $\omega'/m_\pi$, $m_\pi$ being the 
pion mass and $\omega'$ the energy of the real photon. 
This gives rise to the LET and LEX approach: separate out the 
singular part of the amplitude that you can calculate; Taylor expand the 
unknown residual amplitude, thus limiting the unknowns to a (minimal) set 
of constants; relate these constants to the observables and interpret them 
as polarizabilities.
In the next section, I will discuss the general procedure of taking the limit 
of low energy for VCS. 
\section{Low energy theorem for VCS}
\label{LET}
Before proceeding with the low energy expansion, one has to specify the way 
the low energy limit is realized. The problem is twofold: firstly, it has to be 
controlled that when performing the low energy limit all the symmetries of the 
Compton amplitude remain intact; secondly, the Born part that is to be 
separated out is singular precisely at the point where the LEX has to be 
performed. This implies that the choice of the kinematical point for this 
exoansion should be made with care. For instance, for real Compton scattering, 
the low energy limit of the Compton amplitude is well known from classical 
electrodynamics and is given by the constant Thomson term. However, if putting 
only the nucleon in the direct channel on-shell but not the crossed channel 
one, it would be impossible to obtain a constant since a pole cannot be 
cancelled by a regular function. Therefore, unless one goes to the point 
$s=u=M^2$ one would never find the correct low energy limit. 
The point where zero energy limit should be taken is analogous for VCS,
\beqn
s&=&u=M^2\nn\\
\nu&=&0\nn\\
t+Q^2&=&-2(qq')=0
\eeqn
\indent
Note that in Breit frame both initial and final photon energy vanishes at this 
point simultaneously. 
As it was stated in \cite{lexvvcs}, this limit should be realized as 
\beqn
\nu\to0\;\;\;{\rm at\;fixed\;\;}Q^2\;\;{\rm and}\;\; (qq')=0
\eeqn
in order to ensure that i) the path on the Mandelstam plane along which the 
limit is taken lies completely inside the physical region for $s$ or $u$ 
channel process, and ii) this limit can be approached symmetrically either 
from positive ($s$-channel) or negative ($u$-channel) values of $\nu$. 
In Breit frame, this limit corresponds to angle $\theta\to90^\circ$ since 
$\cos\theta((qq')=0)=\frac{\omega}{q}\to0$. Since the value of energy 
should be small enough, $\omega\lesssim m_\pi$, this condition is too 
restrictive on the values of scattering angles for which the above LEX 
prescription is viable. 
To access all the kinematics, one have to relax the condition $(qq')=0$ but 
ensure that $(qq')$ and its first derivative with respect to $\nu$ vanishes at 
$\nu=0$. Vanishing of the first derivative is required by observing that 
$(qq')$ is even under crossing $\nu\to-\nu$.
\begin{figure}[h]
{\includegraphics[height=10cm]{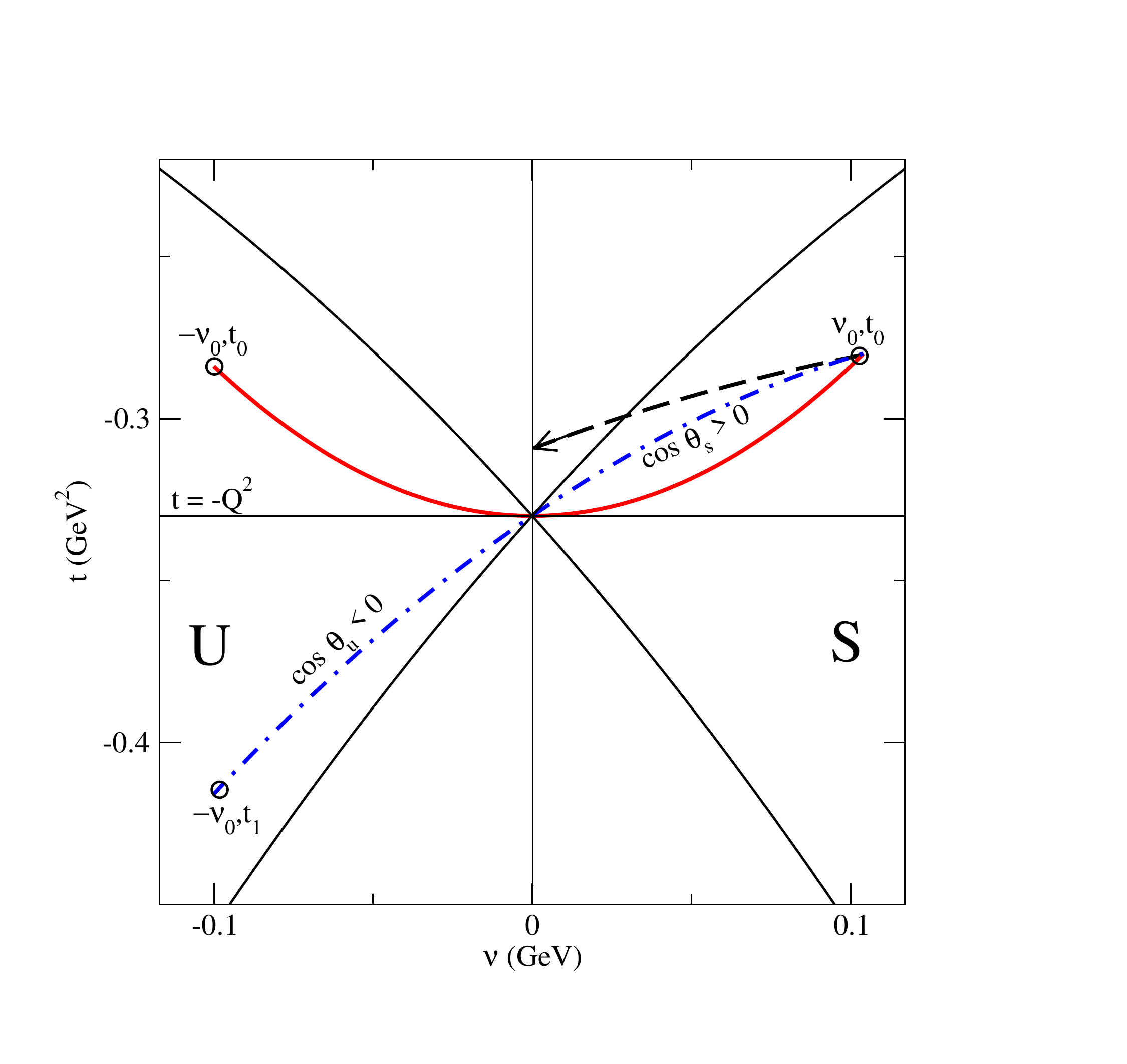}}
\caption{(Color online) The paths on the Mandelstam plane $\nu,t$ for VCS 
with $Q^2=0.33$ 
GeV$^2$, along which the low energy limit can be approached. Starting at the 
VCS measurement kinematics $\nu_0,t_0$, one approaches the low energy limit 
along the crossing-symmetric path of Eq. (\ref{eq:lexpath}) (solid line), 
or at fixed scattering angle (dash-dotted line). Both paths are continued into 
the $u$-channel to illustrate that unlike the crossing-symmetric path, the 
fixed-angle path connects forward 
scattering in the $s$-channel to backward scattering in the $u$-channel. 
For comparison, the low energy limit realization of Guichon et al. 
\cite{guichon} is shown 
by the dashed line that starts at the same point $\nu_0,t_0$. It is seen that 
it connects the VCS measurement at the original value of $Q^2$ to a low energy 
limit of another VCS reaction with a lower value of $Q^2$, that lies outside 
the physical region of the reaction under study.}
\label{fig:lex}
\end{figure}
Therefore, the proposed procedure to perform the low energy limit for VCS is
the following. Assume that a VCS measurement is carried out at the kinematical 
point $\nu_0,t_0$ ($\omega_0=\frac{M\nu_0}{\sqrt{M^2-t_0/4}}$ and 
$|\vec q_0|=\sqrt{\omega_0^2+Q^2}$ accordingly). The 
limit of low energy, $\nu=0,t=-Q^2$ can be approached 
along the path $t(\omega,\omega_0,t_0)$ given by 
\beqn
t(\omega,\omega_0,t_0)+Q^2&=&\frac{\omega^2}{\omega_0^2}(t_0+Q^2)
\;\;\;\;{\rm or}\nn\\
\cos\theta&=&\frac{\omega |\vec q_0|}{\omega_0|\vec q|}\cos\theta_0,
\label{eq:lexpath}
\eeqn
shown in Fig. \ref{fig:lex}.
In the limit of $Q^2=0$ this path reduces simply to $\cos\theta=\cos\theta_0$, 
fixed angle that is used in LEX for RCS. 
There exist more than one functional form that satisfy crossing condition and 
reduce to fixed angle for real Compton scattering. However, they can only 
differ by corrections in powers of $\omega/M$, and are equivalent for LEX. 
For comparison, in \cite{guichon}, the low energy limit is realized at fixed 
scattering angle $\theta_0$ and fixed three-vector magnitude $|\vec q|$. 
Then, along that path the VCS amplitude is decomposed into a series of 
multipoles with definite orbital momentum in the initial and final channel, and 
this series is truncated at the dipole order for the outgoing photon.

For a function that is regular and analytical in the vicinity of the point 
$\nu=0,\;\;(qq')=0$ (as the non-Born VCS amplitude is by construction), that 
point can be approached along any path, and the result should be 
path-independent. 
However, if the function is decomposed into a series in powers of energy 
and truncated at a given order, the path independence can become hard to 
control. The above discussion implies that performing 
the low energy limit at fixed angle for VCS can lead to complicated 
correlations between different terms in the low energy expansion. 
Just because that path on Mandelstam plane is asymmetric, the crossing symmetry 
of the VCS amplitude enforces constraints onto the strength of different 
multipoles, as shown in \cite{andreas}. The multipoles are designed to form a 
basis, order by order in $\omega'$, and existence of such correlations 
indicates the breakdown of the formalism. 
\footnote{Apart from the fact that in \cite{guichon} the low 
energy limit is realized in a crossing-asymmetric way, 
further problems appear: by keeping $|\vec q|$ fixed and varying $\omega'$, 
one actually varies the $Q^2$ which is a Lorentz invariant. Starting at 
$\omega'_0,|\vec q|$ and letting $\omega'\to0$, results in 
$Q^2\to\tilde Q^2$ with $\tilde Q^2\approx Q^2/(1+\omega'_0/M)$. 
So one in practice relates observables 
at one value of $Q^2$ to polarizabilities at a different $Q^2$, as shown in 
Fig. \ref{fig:lex}
The kinematical point of zero energy for $\tilde Q^2$ lies 
outside the physical region for the original VCS process with $Q^2$, then it is 
no surprise that analyticity problems might come up.}
Then, the prescription of Eq. \ref{eq:lexpath} can be seen as a convenient 
choice to incorporate the crossing symmetry of the VCS amplitude to its 
expansion in powers of energy, order by order.
But having the scattering angle depend on 
the energy immediately invalidates the multipole expansion approach to LEX 
since the multipoles and the corresponding harmonics have now to be evaluated 
at $\cos\theta=0$ and higher orders can contribute at the same order as the 
lowest ones.

\section{Expansion of the VCS amplitude in Breit frame and the new set of GP's}
\label{LEX}
In this Section, low energy expansion of the non-Born part of the VCS amplitude 
will be performed. As it was mentioned before, the polarizabilities are in 
general frame dependent quantities since they parametrize Hamiltonian, 
rather than Lagrangean, and they multiply various combinations of the 
electromagnetic field three-vectors. 
This can be easily seen if one considers, for instance the structure
$(P^\nu F'_{\nu\alpha})$ appearing in Eq. (\ref{eq:tensor1}). The electric and 
magnetic fields can be read off the tensors $F$ as 
${F'}^{0i}=\vec {E'}^i$, ${F'}^{ij}=\epsilon^{ijk}\vec {B'}^k$. 
In Breit frame, this structure is purely electric, 
$P^0\vec E\,'$. In the c.m., it contains electric 
and magnetic fields since 
$P=(P+K)-K=(\sqrt{s},\vec0) - ((\omega+\omega')/2, (\vec q+\vec q\,')/2)$, so 
there will be terms $\sim\vec q\times\vec B\,'$ and such. Such terms are higher 
order in $\omega$ for RCS, but are not suppressed for VCS.

I will approach this problem from a slightly different prospective, that of 
effective field theory (EFT). Low energy expansion corresponds to pionless 
EFT, integrating the pion-mediated interactions out of the Lagrangean and 
replacing them with a number of contact interactions that are then organized 
hierarchically in powers of $\omega/m_\pi$. These interactions are 
characterized by the corresponding number of constants that can be related to 
the observables. 
The operator basis of the Lagrangean was already introduced in 
Eqs. (\ref{eq:tensor1},\ref{eq:tensor2}). The corresponding amplitudes can be 
expanded into 
a series in powers of $\nu$ (rather than $\omega$ to keep Lorentz invariance) 
and only the leading order coefficients of this expansion can be kept. 
According to the properties of the amplitudes $F_i$ under crossing, we can 
therefore introduce twelve low energy coefficient functions $f_i(Q^2)$ as 
\beqn
f_i(Q^2)&=&F_i(\nu=0,t=-Q^2,Q^2),\;\;i=1,2,5,6,11,12\nn\\
f_i(Q^2)&=&\frac{1}{\nu}F_i(\nu=0,t=-Q^2,Q^2),\;\;i=3,4,8,10\nn\\
f_i(Q^2)&=&\frac{1}{Q^2}F_i(\nu=0,t=-Q^2,Q^2),\;\;i=7,9
\eeqn
\indent
Pulling out the explicit factors of $\nu$ and $Q^2$ has to be accompanied by a 
redefinition of tensors, $\nu\rho_{3,4,8,10}$ and $Q^2\rho_{7,9}$.
These low energy coefficient functions can now be related to polarizabilities
when specifying a particular reference frame. I will choose the nucleon Breit 
frame and evaluate the basis tensors in that frame to represent them in terms 
of electric and magnetic field three-vectors. 

In RCS, the low energy expansion of the non-Born amplitude starts from 
$\omega^2$ for the spin-independent part, and $\omega^3$ for the spin-dependent 
one. Instead, in VCS this expansion starts at order $\omega$ for both. 
Therefore, one can readily eliminate some of the structures from the tensor 
of Eqs. (\ref{eq:tensor1},\ref{eq:tensor2}) by noticing that any structure 
$\sim Q^2\omega^2$ or $\omega^4$ will contribute at higher order. 

For power counting, one has in Breit frame: $\vec E_L\sim\sqrt{Q^2}$, 
$\vec B\sim|\vec q|$, and $\vec E_T,\vec E',\vec B'\sim\omega$.
The structures to eliminate are 
$\rho_3,4\rho_7-\rho_{11},\rho_8,2\rho_9-\rho_6$ that contribute at order 
$Q^2\omega^2$ (see Appendix B for details) and cannot enter the LEX neither 
for RCS nor for VCS. This, in fact was already observed 
\footnote{Redefining $\rho_{7,9}$ requires redefining the amplitudes 
$F_{6,11}$ as $\tilde F_6=F_6+\frac{1}{2}F_9$, 
$\tilde F_{11}=F_{11}+\frac{1}{4}F_7$. It is these combinations that 
enter the LEX for VCS, and not $F_{6,11}$ \cite{andreas, javcs}.}
\cite{andreas,javcs}. 

As a result, one is left with eight structures that can contribute to LEX at 
lowest order, $\rho_{1,2,4,5,6,10,11,12}$ times the corresponding amplitudes 
$F_{1,2,4,5,6,10,11,12}$. Six of them are relevant for RCS, 
$\rho_{1,2,4,6,10,11}$ while $\rho_{5,12}$ vanish for real photons. 
In turn, for VCS it is $\rho_{4,10}$ that do not contribute to the LEX due to 
the crossing behavior of the respective amplitudes, as found in \cite{andreas}. 
This is the formal origin of the mismatch between the low energy expansions of 
the RCS and VCS. 

The main idea of this work is to {\it assume} that a low-energy reduction of 
the VCS amplitude can be found, that would consist of only six basis structures 
that should be the same for RCS and VCS. 
This amounts in building four linear combinations out of 
$\rho_{4,5,10,12}$ such that two of them give the right limit at low energies, 
whereas the other two should be subleading in LEX. If building such linear 
combinations can be realised in a covariant way and without introducing any 
spurious kinematical singularities, this would be the solution to the problem. 
After a little algebra reported in Appendix B, such combinations are
\beqn
\phi_1&=&2\rho_5\,-\,\nu\rho_{10}\nn\\
\phi_2&=&2\nu\rho_5\,+\,Q^2\rho_{10}\nn\\
\phi_3&=&4M\rho_5-\tilde\rho_{12}-\nu\rho_{4}\nn\\
\phi_4&=&-\nu\tilde\rho_{12}+Q^2\rho_{4},
\eeqn
with combinations $\phi_{2,4}$ subleading (i.e., $\sim\omega^2 Q^2$) by 
construction. The combination $\tilde\rho_{12}=\rho_{12}+\frac{Q^2}{4P^2}\rho_2-\frac{Q^2}{8M}(\rho_{11}-4\rho_5)$ was introduced in order to localize the structure of interest, $i(\vec\sigma\vec B')(\vec q\vec E)$.
\indent
To ensure that the above linear combinations are adequate, one should control 
that the determinant of such transformation is non-zero. In fact, it is 
\beqn
det\left(
\begin{array}{cccc}0&2&-\nu&0\\
0&2\nu&Q^2&0\\
-\nu&4M&0&-1\\
Q^2&0&0&-\nu
\end{array}\right)
=2(\nu^2+Q^2)^2>0
\eeqn
\indent
The new amplitudes $\delta_{1,2,3,4}(\nu,t,Q^2)$ are obtained from the old 
amplitudes as
\beqn
\delta_1&=&\frac{Q^2\tilde F_5-2\nu F_{10}}{2(\nu^2+Q^2)}
+\frac{2M\nu Q^2F_4+2MQ^4F_{12}}{(\nu^2+Q^2)^2}
\nn\\
\delta_2&=&\frac{\nu \tilde F_5+2F_{10}}{2(\nu^2+Q^2)}
+\frac{2M\nu^2F_4+2M\nu Q^2F_{12}}{(\nu^2+Q^2)^2}
\nn\\
\delta_3&=&\frac{-\nu F_4-Q^2 F_{12}}{(\nu^2+Q^2)}\nn\\
\delta_4&=&\frac{F_4-\nu F_{12}}{(\nu^2+Q^2)}
\eeqn
where I kept the full amplitudes $F_{4,5,10,12}$ at arbitrary values of their 
arguments $\nu,t,Q^2$, and $\tilde F_5\equiv F_5+F_7+4F_{11}$ is the 
combination of invariant amplitudes that arises due to combining basis tensors 
in Eq.(\ref{eq:8basis}). 
While we know that the old amplitudes are free of kinematical singularities and 
constraints, we need to make sure that also the new amplitudes possess these 
properties, before we can use them for LEX.

I will now examine the low energy limit of these amplitudes for 
RCS and for VCS, and {\it require} that this limit is continuous at $Q^2=0$, 
according to the general assumption made earlier in this section. 
I consider first $\delta_1$ and make use of the low energy 
coefficient functions $f_i(Q^2)$ introduced earlier. 
\beqn
\delta_1(\nu,t,Q^2=0)
&=&-f_{10}(0),\\
\delta_1(\nu=0,t=-Q^2,Q^2)&=&\frac{1}{2}[\tilde f_5(Q^2)+4Mf_{12}(Q^2)],\nn
\eeqn
with $\tilde f_5\equiv f_5+Q^2f_7+4f_{11}$.  
Requiring that these two limits commute leads to imposing a relation:
\beqn
f_5(0)+4f_{11}(0)+4Mf_{12}(0)&=&-2f_{10}(0),
\label{eq:f5}
\eeqn
\indent
Similarly, one obtains for $\delta_3$
\beqn
\delta_3(\nu,t,Q^2=0)&=&-f_{4}(0),\nn\\
\delta_3(\nu=0,t=-Q^2,Q^2)&=&-f_{12}(Q^2),
\eeqn
and again requiring the two limits to commute leads to another relation,
\beqn
f_{12}(0)&=&f_4(0)
\label{eq:f12}
\eeqn
\indent
These relations also ensure that the low energy limit of the subleading 
amplitudes, $\delta_{2,4}$ is continuous. Thus, given the asymptotics of the 
corresponding tensors $\sim\omega^2 Q^2$, these amplitudes can be eliminated 
from the low-energy effective Lagrangean. 

Summarizing the procedure described above, 
the EFT basis for non-Born part of low energy real and virtual Compton 
scattering has six independent structures,
\begin{widetext}
\beqn
T_{VCS}^{EFT}&=&-\frac{1}{2}f_1(Q^2)F^{\mu\nu}F'_{\mu\nu}
-4f_2(Q^2)\left(P_\mu F^{\mu\alpha}\right)\left(P^\nu F'_{\nu\alpha}\right)
\label{eq:EFT}\\
&+&\frac{1}{2}(f_5(Q^2)+Q^2f_7(Q^2)+4f_{11}(Q^2))
\left[\left(q'^\mu F'_{\mu\alpha}\tilde F^{\alpha\beta}
-q^\mu F_{\mu\alpha}\tilde F'^{\alpha\beta}
\right)i\gamma_5\gamma_\beta
+\frac{Pq}{M}F^{\mu\alpha}{F'}_\mu^\beta i\sigma_{\alpha\beta}\right]
\nn\\
&+&(2f_6(Q^2)+Q^2f_9(Q^2))\left[
\left(P_\alpha q_\beta F'^{\alpha\beta}\right)F^{\mu\nu}
-\left(P_\alpha q'_\beta F^{\alpha\beta}\right)F'^{\mu\nu}
\right]i\sigma_{\mu\nu}\nn\\
&+&\frac{1}{2}\left(Q^2f_7(Q^2)+4f_{11}(Q^2)+\frac{Q^2}{2M}f_{12}(Q^2)\right)
\Delta^\mu\left[
F'_{\mu\alpha}\tilde F^{\alpha\beta}+F_{\mu\alpha}\tilde F'^{\alpha\beta}
\right]i\gamma_5\gamma_\beta\nn\\
&+&f_{12}(Q^2)\left[
\left(P_\alpha q'_\beta F'^{\alpha\beta}\right)F^{\mu\nu}i\sigma_{\mu\nu}
-\left(P_\alpha q_\beta F^{\alpha\beta}\right)F'^{\mu\nu}i\sigma_{\mu\nu}
+\frac{2Pq}{M}P^\mu
\left(F'_{\mu\alpha}\tilde F^{\alpha\beta}-F_{\mu\alpha}\tilde F'^{\alpha\beta}
\right)i\gamma_5\gamma_\beta
\right]\nn
\eeqn
\end{widetext}
and this low energy reduction of the VCS amplitude is continuous in the limit 
$Q^2\to0$. The original numeration of Ref. \cite{andreas} 
of the amplitudes is kept to avoid any confusion. The six low energy 
constants (for fixed $Q^2$) completely describe the effects of the proton 
structure on Compton scattering with real and virtual photons and at leading 
order in low energy expansion. The form of the above tensor suggests that the 
fact that the low energy limits for RCS and VCS were found in previous 
studies to not match with one another, could be attributed to the particular 
way that was used in the literature to 
construct the VCS basis starting from the RCS one. It was done by adding 
structures that explicitly vanish for real photon, and it resulted in only 
partial overlap of the low energy reduction of the VCS basis with the RCS one.

Next, I will introduce the (generalized) Breit frame polarizabilities following 
the approach of Ragusa \cite{ragusa} who complemented the LEX of the 
spin-independent part of Low \cite{low} by introducing 
the four spin-dependent polarizabilities in Breit frame, as well.
While in Ref. \cite{ragusa}, the Compton 
amplitude was written in terms of polarization vectors of the photons in order 
to have an explicit power counting in photon energy, I will 
rather write it in terms of electromagnetic fields. In this way, 
the power counting and 
polarization content (longitudinal or transverse) is implicit, but the 
generality of this description (i.e., RCS and VCS) and the 
analogy with the classical electromagnetic 
polarizabilities become more transparent. 

Following the standard conventions for the Compton amplitude, 
and correcting for Breit kinematics with virtual initial photon, I obtain 
the natural generalization of the Ragusa's LEX:
\begin{widetext}
\beqn
\frac{1}{2P^0}T_{LEX}^{NB}&=&
4\pi\alpha(Q^2)\vec E\cdot\vec E'\chi^\dagger\chi
+4\pi\beta(Q^2)\vec B\cdot\vec B'\chi^\dagger\chi\nn\\
&+&4\pi(\gamma_1(Q^2)-\gamma_2(Q^2)-2\gamma_4(Q^2))\frac{1}{2}
\left[
\vec E\times[\vec q\,'\times\vec B\,']-\vec E\,'\times[\vec q\times\vec B]
\right]\chi^\dagger i\vec\sigma\chi\nn\\
&+&4\pi\gamma_2(Q^2)\frac{1}{2}
\left[
\vec q\,'\times[\vec E\,'\times\vec B]-\vec q\times[\vec E\times\vec B\,']
\right]\chi^\dagger i\vec\sigma\chi\nn\\
&+&4\pi(\frac{1}{2}\gamma_2(Q^2)+\gamma_3(Q^2)+\gamma_4(Q^2))
\left[
\vec B(\vec q\vec E\,')-\vec B\,'(\vec q\,'\vec E)
\right]\chi^\dagger i\vec\sigma\chi\nn\\
&-&4\pi(\frac{1}{2}\gamma_2(Q^2)+\gamma_4(Q^2))
\chi^\dagger i\vec\sigma\vec\Delta\chi
(\vec E\,'\cdot\vec B+\vec E\cdot\vec B\,'),
\label{eq:lex}
\eeqn
\end{widetext}
where $\alpha(Q^2),\beta(Q^2)$, and $\gamma_i(Q^2)$, $i=1,2,3,4$ 
are the generalized polarizabilities for VCS, and the notation is used since 
they reduce to the polarizabiilities of real Compton scattering at $Q^2=0$. 
Note that the operators that multiply the 
polarizabilities are more general than those of Ragusa \cite{ragusa} since 
they should also 
incorporate the virtual incoming photon. In particular, it can be noticed 
that the original structures of Ref. \cite{ragusa} that multiply $\gamma_2$ 
and $\gamma_4$ are not linearly-independent for VCS. I choose an appropriate 
linear combination of the two to accompany $\gamma_2$, and the remaining 
structure is chosen to coincide with that arising due to the physical 
contribution of the $\pi^0$ exchange in the $t$-channel. One can easily 
derive the relations between the basis structure listed above and the original 
ones of Ragusa, by diagonalizing Eq. (\ref{eq:lex}) with respect to 
$\gamma$'s.

An important feature of the basis of Eq. (\ref{eq:lex}) is that no distinction 
is made for 
transverse or longitudinal polarization of the virtual photon, as for instance 
in \cite{scherer}. In that reference, two different electric polarizabilities 
were introduced, $\alpha_L$ and $\alpha_T$, and it was then shown in a model 
that $\alpha_L$ is dominant. 
A similar result is obtain here, with the only difference that while in 
\cite{scherer} this dominance of $\alpha_L$ over $\alpha_T$ is realized as 
function of $Q^2$, in the present work the dominance is in $\omega$, i.e. due 
to the neglect of terms $\sim\omega^2 Q^2$ that go beyond the LEX precision. 
Phenomenologically, it is important to realize that even if two distinct 
electric polarizabilities may be introduced in a special frame, there is no 
practical way to determine them both in LEX formalism.

In the remainder of this section, I list the relations between the generalized 
polarizabilities and the Lorentz invariants $f_i(Q^2)$ listed earlier, and 
obtain the two missing relations between $\gamma_i$'s and the GP's of Guichon 
et al.
\begin{widetext}
\beqn
4\pi\beta(Q^2)&=&-f_1(Q^2)\nn\\
4\pi\alpha(Q^2)&=&f_1(Q^2)+4P^2f_2(Q^2)
+Q^2\left[2f_6(Q^2)+Q^2f_9(Q^2)-f_{12}(Q^2)\right]\nn\\
4\pi\gamma_1(Q^2)&=&\frac{4M}{P^0}
\left[\frac{Q^2}{4}f_7(Q^2)+f_{11}(Q^2)\right]-
\left[f_5(Q^2)+Q^2f_7(Q^2)+4f_{11}(Q^2)\right]\nn\\
4\pi\gamma_2(Q^2)&=&
\left[f_5(Q^2)+Q^2f_7(Q^2)+4f_{11}(Q^2)+4Mf_{12}(Q^2)\right]\nn\\
4\pi\gamma_3(Q^2)&=&
-\frac{2M}{P^0}
\left[\frac{Q^2}{4}f_7(Q^2)+f_{11}(Q^2)\right]
-2M\left[2f_6(Q^2)+Q^2f_9(Q^2)\right]\nn\\
4\pi(\gamma_2(Q^2)+\gamma_4(Q^2))&=&
\frac{2M}{P^0}
\left[\frac{Q^2}{4}f_7(Q^2)+f_{11}(Q^2)\right]
\label{eq:GPs}
\eeqn
\end{widetext}
These relations represent the definition of the GP's for finite $Q^2$, and 
the correct limit at $Q^2=0$ is ensured by the relations of Eqs. 
(\ref{eq:f5},\ref{eq:f12}). While some of the relations only contain Lorentz 
scalars, the presence of a factor $\frac{M}{P^0}$ in the others means that 
there is a fundamental frame dependence in expanding VCS observables in powers 
of energy times polarizabilities, and this dependence is expressed in terms of 
recoil corrections $\sim\frac{Q^2}{M^2}$.

In the above relations, the new GP's are given in 
terms of $f_5,f_{12}$ unlike the old GP's 
$P^{(01,01)1},P^{(11,11)1}\sim Q^2f_5,Q^2f_{12}$ \cite{javcs} that vanished 
for real photons. Then, it is the slope of these two GP's at $Q^2=0$ 
that should match the RCS polarizabilities, and not their values at $Q^2=0$,
\beqn
\frac{4\pi}{e^2}\gamma_1(0)&=&-6M\frac{d}{dQ^2}P^{(11,11)1}(0)\nn\\
\frac{4\pi}{e^2}\gamma_2(0)&=&6M\frac{d}{dQ^2}P^{(01,01)1}(0),
\label{eq:gamma12-GPs}
\eeqn
where relations between $f_i$'s and GP's \cite{javcs} were used along with the 
results of Eq. (\ref{eq:GPs}).
Eqs. (\ref{eq:f5},\ref{eq:f12},\ref{eq:gamma12-GPs}) state 
model-independent relations that are only based on general properties 
of the VCS amplitude, plus the {\it assumption} of its analyticity at $Q^2=0$. 
This assumption is worth checking in models, and chiral pereturbation theory 
seems to be the perfect tool to study VCS at very low energy and $Q^2$. 
Calculations of the low energy VCS amplitude exist in the linear $\sigma$-model 
\cite{metz}, heavy-baryon ChPT \cite{hemmert,kao}, effective Lagrangian model 
\cite{korchin}, non-relativistic constituent quark model \cite{barbara} 
(for a recent review, see \cite{walcher}). 
Unfortunately, none of the above references include both real and virtual 
Compton scattering within the same formalism, and moreover frame dependence may 
be crucial since it introduces corrections $\sim Q^2/M^2$ that alter the slope 
of the GP's that have to be computed. Therefore, the two relations 
either should be checked on the level of the invariant amplitudes, or both 
sides of the two equalities should be evaluated within the same model. 
The author leaves this for an upcoming work.

\section{VCS observables}
\label{sec:obs}
\begin{figure}[h]
{\includegraphics[height=7.5cm]{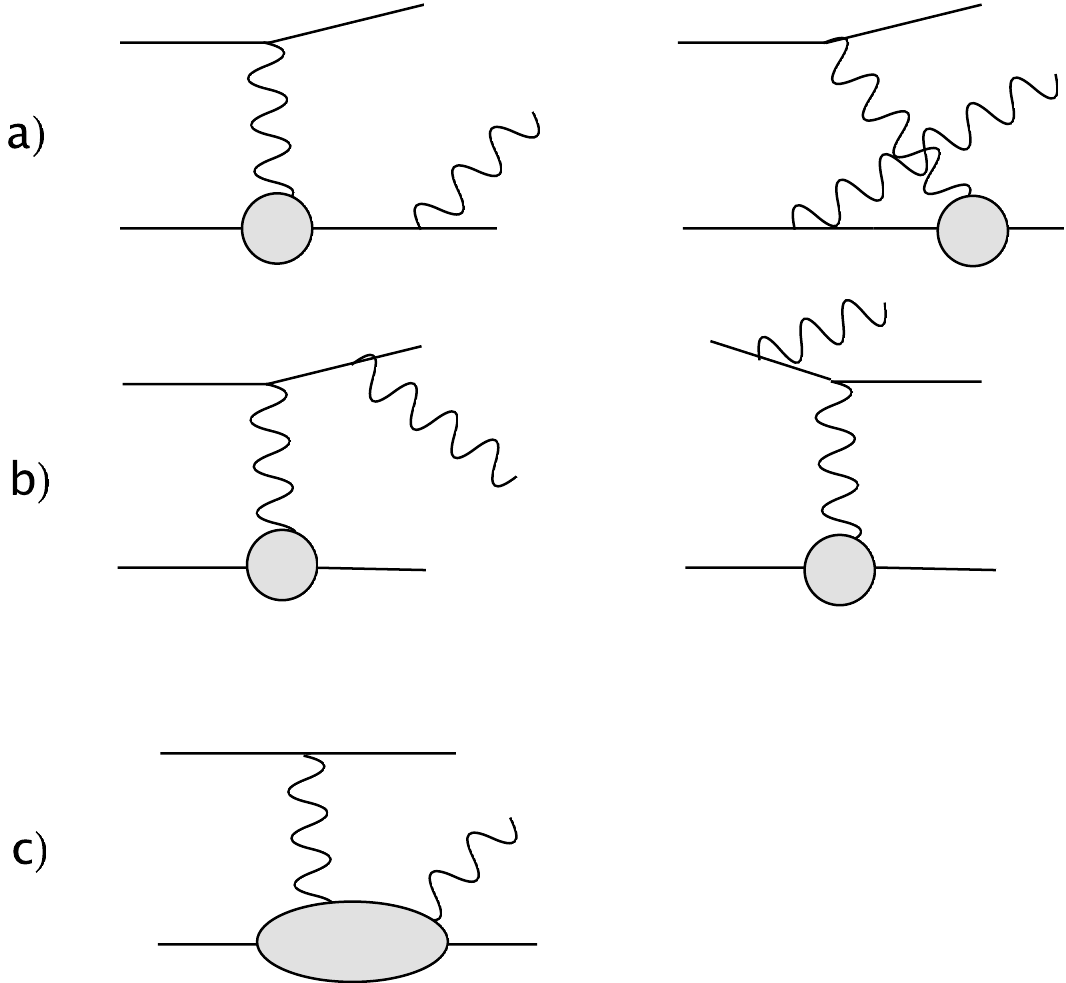}}
\caption{The different contributions to the scattering process 
$e+p\to e+p+\gamma$: FVCS Born (case a), Bethe-Heitler (case b), and non-Born 
FVCS (case c). For a) and b), the blob denotes the elastic nucleon form 
factors.}
\label{fig:ep-epga}
\end{figure}

Since the set of GP's introduced in the present work is different from those of 
Guichon et al., I will consider the effect of the new GP's on the observables, 
and since LEX was performed in Breit frame, the same frame will be used in 
this section, too. I will repeat the main steps that were done in 
\cite{guichon} for c.m. kinematics. 
In the process $e+p\to e+p+\gamma$, the real photon can be emitted from one 
of the electron legs (Bethe-Heitler, shown in Fig. \ref{fig:ep-epga},b), 
from a local coupling to one of the nucleon legs 
(Born, Fig. \ref{fig:ep-epga},a) or can originate from a non-local two-photon 
interaction (non-Born, Fig. \ref{fig:ep-epga},c), 
\beqn
T_{ep\to ep\gamma}&=&T_{BH}+T_{FVCS}^B+T_{FVCS}^{NB}
\eeqn
\indent
Bethe-Heitler amplitude is given by
\begin{widetext}
\beqn
T_{BH}&=&-\frac{e^3}{t}\bar N(p')\Gamma_\mu(\Delta)N(p)
\bar u(k')
\left[\frac{\gamma^\nu(\sk'+\sq'+m_e)\gamma^\mu}{(k'+q')^2-m_e^2}+
\frac{\gamma^\mu(\sk-\sq'+m_e)\gamma^\nu}{(k-q')^2-m_e^2}\right]u(k)
{\varepsilon'}^*_\nu
\eeqn
\end{widetext}
where $m_e$ is the electron mass, and ${\varepsilon'}_\nu$ stands for the 
polarization vector of the outgoing real photon.
Similarly, Born contribution is given by
\begin{widetext}
\beqn
T^B_{FVCS}&=&-\frac{e^3}{Q^2}\bar u(k')\gamma_\mu u (k)
\bar N(p')
\left[\frac{\Gamma^\nu(q')(\sp'+\sq'+M)\Gamma^\mu(q)}{(p'+q')^2-M^2}+
\frac{\Gamma^\mu(q)(\sp-\sq'+M)\Gamma^\nu(q')}{(p-q')^2-M^2}\right]N(p)
{\varepsilon'}^*_\nu
\eeqn
\end{widetext}
These two amplitudes can also be expanded into a series in powers of the 
outgoing photon energy. The expansion strarts with $\nu^{-1}$ since both 
amplitude diverge at zero energy. 
The regular part of the VCS amplitude is embedded into the full amplitude in 
a similar manner,
\beqn
T_{FVCS}&=&\frac{e}{Q^2}\bar u(k')\gamma_\mu u (k)
\sum_{i=1}^{12}\bar N(p')\rho_i^{\mu\nu}F_i(\nu,t,Q^2)N(p)
{\varepsilon'}^*_\nu\nn\\
\eeqn
and its expansion in energy starts at $\nu^1$. 
The differential $(e,e'\gamma)$ cross section is related to the squared 
amplitude,
\begin{widetext}
\beqn
d^5\sigma\,\sim\,
|T_{BH}+T_{FVCS}^B+T_{FVCS}^{NB}|^2
&=&|T_{BH}+T_{FVCS}^B|^2
+\left[(T_{BH}+T_{FVCS}^B)^*T_{FVCS}^{NB}
+(T_{BH}+T_{FVCS}^B)T_{FVCS}^{NB*}\right]\nn\\
&+&|T_{FVCS}^{NB}|^2
\eeqn
\end{widetext}
and it allows for an expansion in powers of $\omega$, 
\beqn
&&|T_{BH}+T_{FVCS}^B|^2=\frac{a^{BH+B}_{-2}}{\omega^2}
+\frac{a^{BH+B}_{-1}}{\omega}+a^{BH+B}_{0}+O(\omega)\nn\\
&&\left[(T_{BH}+T_{FVCS}^B)^*T_{FVCS}^{NB}
+(T_{BH}+T_{FVCS}^B)T_{FVCS}^{NB*}\right]\nn\\
&&=a^{GP}_0+O(\omega)
\nn\\
&&|T_{FVCS}^{NB}|^2=O(\omega^2).
\eeqn
\indent
Correspondingly, it was proposed in \cite{guichon} to extract the GP's from 
the discrepancy of the measured cross section, on one hand, and the 
Bethe-Heitler plus Born cross section that can be calculated, on the other 
hand. This amounts in calculating the coefficient 
\beqn
a^{GP}_0&=&(T^{-1}_{BH}+T_{FVCS}^{B,-1})^*T_{FVCS,1}^{NB}\nn\\
&+&(T^{-1}_{BH}+T_{FVCS}^{B,-1})T_{FVCS,1}^{NB*}
\eeqn
of the interference between the leading $\sim1/\omega$ terms of the BH+B part 
and the $\sim\omega$ term of the non-Born amplitude that is parametrized in
terms of GP's. 
The (model-independent) divergent parts are given by
\beqn
T_{BH}^{-1}&=&-\frac{e^3}{t}\bar u(p')\Gamma_\mu(\Delta)u(p)
\bar u(k')\gamma^\mu u(k)\nn\\
&\times&\left[\frac{{k'}^\nu}{(k'q')}-\frac{{k}^\nu}{(kq')}\right]
{\varepsilon'}^*_\nu\nn\\
T_{FVCS}^{B,-1}&=&-\frac{e^3}{Q^2}\bar u(p')\Gamma_\mu(q)u(p)
\bar u(k')\gamma^\mu u(k)\nn\\
&\times&\left[\frac{{p'}^\nu}{(p'q')}-\frac{{p}^\nu}{(pq')}\right]
{\varepsilon'}^*_\nu,
\eeqn
and the leading term of the non-Born part is given by 
\beqn
T_{FVCS}^{NB,1}=\frac{e}{Q^2}
\bar u(k')\gamma_\mu u(k)
\bar u(p')\sum_i\rho_i^{\mu\nu}f_i(Q^2)u(p){\varepsilon'}^*_\nu\nn\\
\eeqn
with $f_i(Q^2)$ related to the GP's as in Eq.(\ref{eq:GPs}). 
Note that for the leading term, $t$ can be substituted with $-Q^2$ since 
$t+Q^2=-2(qq')\sim\omega$.\\ 
\indent
I will only deal here with the unpolarized case. 
After some algebra, details of which can be found in the Appendix, the 
coefficient $a_0^{GP}$ can be represented in the familiar form 
(cf. \cite{guichon})
\beqn
a_0^{GP}&=&\frac{8Me^4}{1-\epsilon}
\left[v_1(\epsilon P_{LL}-P_{TT})+v_2P_{LT}
\right],
\eeqn
in terms of three structure functions that are related to the GP's as 
\beqn
P_{LL}&=&G_E\left[4\pi\alpha+\frac{Q^2}{2M}
4\pi(\gamma_1+\gamma_2+2\gamma_3)\right]
\nn\\
P_{TT}&=&\frac{Q^2}{4M}G_M 4\pi(\gamma_1+2\gamma_3)
\nn\\
P_{LT}&=&-G_E 4\pi\beta+\frac{Q^2}{4M}G_M 4\pi\gamma_2,
\label{eq:strfun}
\eeqn
with the dependence of the GP's and form factors on $Q^2$ suppressed for 
shortness. $v_1$ and $v_2$ are shrothands for somewhat lengthy kinematical 
factors that are listed in the Appendix, and 
$\epsilon$  denotes the usual virtual photon polarization parameter, 
and is also given in the Appendix.
As already known in the literature, three independent structure functions can 
be measured in the unpolarized VCS experiment at low energies. Now, they all 
are expressed through the polarizabilities that are a direct generalization of 
those of RCS. 
Comparing Eq. (\ref{eq:strfun}) to the original results of \cite{guichon}, one 
notices the presence of the spin-dependent GP's in the structure function 
$P_{LL}$. The origin for that contribution is in terms 
$\sim {F'}^{0\nu}i\sigma_{0\nu}$ that give the spin-independent term 
$\sim-\vec\Delta\vec E\,'\chi^\dagger\chi$. This difference may be 
attributed to the use of the Breit kinematics instead of the c.m. kinematics 
used in \cite{guichon}. 
These new terms in $P_{LL}$ are expected to be less important for low and 
moderate values of $Q^2$, as in the kinematics of Mainz \cite{vcsmami} and 
MIT-Bates \cite{vcsbates} experiments, 
but might affect the extracted values of $\alpha(Q^2)$ and $\beta(Q^2)$ for 
the kinematics of the JLab experiments \cite{vcsjlab}. \\

\section{Summary}
To summarize, I considered the virtual Compton scattering process at low 
energy of the outgoing real photon. I formulated the low energy theorem for 
that reaction, and the present formulation is realized in an explicitly 
crossing-symmetric way that is an improvement with respect to the previous 
formalism \cite{guichon}. I demonstrated that for virtual photons, the 
requirement of crossing symmetry of the VCS amplitude makes the LEX in terms 
of multipole expansion too complicated, if at all viable since it introduces 
relations upon different multipole transitions that are supposed to form a 
basis at leading order in that expansion. Instead, I proposed a different 
approach based on a Lorentz covariant EFT description of VCS at low energies. 
Within this approach, it was possible to define the low energy limit in 
a continuous way, with respect to the virtuality of the initial 
photon, and the same six structures and associated with them low energy 
coefficient functions fully describe Compton scattering at low energies with 
real and virtual photons. These six low energy constants (for fixed $Q^2$) can 
be interpreted as polarizabilities only when going to a specific reference 
frame. I chose Breit frame since it treats the initial and final photons in a 
symmetric way, and crossing symmetry and power counting are realized in 
a simple manner in that frame. Using the classical notion of the polarizability 
and working in a framework closely related to that of Ref. \cite{ragusa} 
where the complete set of nucleon polarizabilities was introduced for RCS, 
I obtained the new set of the six generalized polarizabilities. These new GP's 
are defined such as to reduce to the polarizabilities of RCS for real 
initial photon, the feature that was missing in the formalism of 
\cite{guichon}. The continuous limit at $Q^2=0$ imposes two relations between 
the values of four invariant amplitudes for VCS at the real photon point, 
leading to two relations of $\gamma_{1,2}$ to the slope of two GP's of 
\cite{guichon} in that kinematical point. These
two relations should be checked in models, most notably within chiral 
perturbation theory in its relativistic or heavy baryon form.
I also computed the contribution of the GP's to the unpolarized VCS cross 
section in Breit frame. While confirming the general structure of this 
contribution as reported in \cite{guichon,andreas}, I found that the structure 
function $P_{LL}$ has a contribution from spin-dependent GP's that is not 
present in the analysis of \cite{guichon}. 
\begin{acknowledgments}
The author is grateful to J.T. Londergan and B. Holstein for stimulating 
discussions. This work was supported by the US National Science Foundation 
under grant PHY 0555232.
\end{acknowledgments}
\begin{appendix}
\section{VCS kinematics, polarization vectors and nucleon spinors in 
Breit frame}
\label{app:a}
I use the standard definition of the nucleon Breit frame,
\beqn
P^\mu&=&(P^0,\vec0)\nn\\
\Delta^\mu&=&(0,0,0,|\vec\Delta|)
\eeqn
with $t=-\vec\Delta^2$, and the energy of the initial (final) nucleon $E(E')$ 
are equal to $E=E'=P^0=\sqrt{M^2-\frac{t}{4}}$.
I use $\Delta\equiv|\vec\Delta|$ and 
$q\equiv|\vec q|$ in the following. The photons' momenta in this frame are 
given by 
\beqn
q^\mu&=&(\omega,q\cos\alpha,0,q\sin\alpha)\nn\\
q'^\mu&=&\omega(1,\cos\beta,0,-\sin\beta)\nn\\
\sin\beta&=&\frac{t+Q^2}{2\omega\Delta}\nn\\
\cos\alpha&=&\frac{\omega}{q}\cos\beta
\eeqn
\indent
In the following, 
I will also use the photon kinematics in the limit of very small photon energy, 
\beqn
q^\mu&\approx&\Delta^\mu=(0,0,0,\Delta)\nn\\
q'^\mu&=&\omega(1,\sin\theta,0,\cos\theta)
\eeqn
with $\theta=\frac{\pi}{2}+\beta$.
The polarization vectors for transverse photons (also notation 
$\vec\varepsilon_T$ is used in the text) are 
\beqn
\vec\varepsilon_{\lambda=\pm}^\mu(\vec q)&=&-\frac{\lambda}{\sqrt{2}}
(0,\sin\alpha,i\lambda,-\cos\alpha)\nn\\
\vec\varepsilon_{\lambda'=\pm}^{'\mu}(\vec q')&=&-\frac{\lambda'}{\sqrt{2}}
(0,-\sin\beta,i\lambda',-\cos\beta).
\eeqn

For the longitudinal polarization of the virtual photon, one has
\beqn
\vec\varepsilon_{\lambda=0}^\mu(\vec q)&=&\frac{1}{\sqrt{Q^2}}
(q,\omega\hat q)
\eeqn
with $\hat q$ the unit vector in the direction of the virtual photon's 
three-momentum. One can identically rewrite it as
\beqn
\vec\varepsilon_{\lambda=0}^\mu(\vec q)
\;=\;\frac{q}{\omega\sqrt{Q^2}}q^\mu-\frac{\sqrt{Q^2}}{\omega}(0,\hat q)
\eeqn
and using gauge invariance of the VCS tensor,
$q_\mu T^{\mu\nu}=0$, one can verify that the first term does not contribute. 
The longitudinal polarization vector (denoted as $\vec\varepsilon_L$ in the 
body of the article) of the initial photon is used in the form
\beqn
\vec\varepsilon_{\lambda=0}^\mu(\vec q)
\;=\;-\frac{\sqrt{Q^2}}{\omega}(0,\hat q).
\eeqn

For nucleon spinors describe a Dirac particle with the three-vector 
$\vec p$, mass $M$ and energy $E=\sqrt{\vec p^2+M^2}$, one has
\beqn
u(\vec p)&=&\sqrt{E+M}\left[
\begin{array}{c}
\chi\\\frac{\vec\sigma\vec p}{E+M}\chi
\end{array}
\right],\nn\\
u(\vec p')&=&\sqrt{E+M}\left[
\begin{array}{c}
\chi\\\frac{\vec\sigma\vec p'}{E+M}\chi
\end{array}
\right],
\eeqn
and due to Breit kinematics one has 
$\vec\sigma\vec p'=-\vec\sigma\vec p=\frac{1}{2}\vec\sigma\vec\Delta$.
Pauli spinors $\chi$ are taken to correspond to a definite $z$-projection of 
the nucleon spin both for the initial and final nucleons, 
\beqn
\chi_{+1/2}=\left(\begin{array}{c}1\\0\end{array}\right)&&
\chi_{-1/2}=\left(\begin{array}{c}0\\1\end{array}\right)
\eeqn
\indent
With these definitions, one finds the following useful relations:
\beqn
\bar u(p')u(p)&=&2E\chi^\dagger\chi\nn\\
\bar u(p')\vec\gamma\gamma_5u(p)&=&
2E\chi^\dagger\left[\vec\sigma-\frac{E-M}{E}\vec\sigma_z\right]\chi\nn\\
\bar u(p')\gamma^0\gamma_5u(p)&=&0\nn\\
\bar u(p')i\sigma^{ij}u(p)&=&i\epsilon^{ijk}2M
\chi^\dagger\left[\vec\sigma^k+\frac{E-M}{M}\vec\sigma_z^k\right]\chi\nn\\
\bar u(p')i\sigma^{0i}u(p)&=&\vec\Delta^i\chi^\dagger\chi
\eeqn
\indent
I use conventions 
$\gamma_5=i\gamma^0\gamma^1\gamma^2\gamma^3=\left(\begin{array}{cc}0&1\\1&0\end{array}\right)$, and $\epsilon_{0123}=+1$.

\section{Power counting in Breit frame and low energy reduction of the VCS 
invariant basis}
\label{app:a1}
For power counting, one has in Breit frame
\beqn 
\vec E'&=&\vec E'_T=ie\omega\vec\varepsilon'\sim\omega\nn\\
\vec B'&=&ie[\vec q'\times\vec\varepsilon']\sim\omega\nn\\
\vec E_T&=&ie\omega\vec\varepsilon_T\sim\omega\nn\\
\vec E_L&=&ie\omega\vec\varepsilon_L\sim\sqrt{Q^2}\nn\\
\vec B&=&ie[\vec q\times\vec\varepsilon_T]\sim|\vec q|,
\eeqn
and I refer the reader to the Appendix A for explicit expressions of the 
polarization vectors. The structures to eliminate are 
\beqn
&&\nu\rho_3=\frac{2}{M}
\left[-Q^2g_{\alpha\beta}-q_\alpha q_\beta\right]
\left(P_\mu F^{\mu\alpha}\right)\left(P_\nu F'^{\nu\beta}\right)\nn\\
&&Q^2(\rho_7-\frac{1}{4}\rho_{11})=Q^2
(q'^\mu F_{\mu\alpha})\tilde F'^{\alpha\beta}i\gamma_5\gamma_\beta\nn\\
&&\nu\rho_8=
\nu\frac{1}{2}\left(q_\mu q_\nu+q'_\mu q'_\nu\right)
F^{\mu\alpha}F'^{\nu\beta}i\sigma_{\alpha\beta}\nn\\
&&-\nu\frac{1}{4}q_\alpha q'_\beta
\left[F^{\alpha\beta}F'^{\mu\nu}+F'^{\alpha\beta}F^{\mu\nu}\right]
i\sigma_{\mu\nu}\nn\\
&&Q^2(\rho_9-\frac{1}{2}\rho_{6})=
2Q^2\left(P_\alpha q'_\beta F^{\alpha\beta}\right)F'^{\mu\nu}i\sigma_{\mu\nu}
\nn\\
&&+Q^2\frac{(qq')}{4M}(\rho_{11}-4\rho_5)
\eeqn
\indent
All the structures listed above contribute at order $Q^2\omega^2$ and cannot 
enter the LEX neither for RCS nor for VCS. 
\footnote{The only tensor for which it is not obvious right away is $\rho_3$. 
Explicit evaluation in terms of $\vec E$ fields gives
$\nu\rho_3\sim Q^2(\vec E'\vec E)-(\vec q\vec E')(\vec q\vec E)= Q^2(\vec E'\vec E_T)-\omega^2(\vec E'\vec E_L)$.}

As a result, one is left with eight structures that can contribute to LEX at 
lowest order, $\rho_{1,2,4,5,6,10,11,12}$ times the corresponding amplitudes 
$F_{1,2,4,5,6,10,11,12}$. Six of them are relevant for RCS, 
$\rho_{1,2,4,6,10,11}$ while $\rho_{5,12}$ vanish for real photons. 
In turn, for VCS it is $\rho_{4,10}$ that do not contribute to the LEX due to 
the crossing behavior of the respective amplitudes, as found in \cite{andreas}. 
This is the formal origin of the mismatch between the low energy expansions of 
the RCS and VCS. 
I will review the situation in detail by rewriting the eight tensors in terms 
of electromagnetic fields in Breit frame following in grand line Refs. 
\cite{ragusa,scherer}.
Understanding in the following the tensors $\rho_i$ cast between the 
initial and final nucleon (Pauli) spinors, these expressions read
\beqn
&&\rho_1=2P^0[(\vec E\vec E')-(\vec B\vec B')]\label{eq:8basis}\\
&&\rho_2=8(P^0)^3(\vec E\vec E')\nn\\
&&\nu\rho_4=4\frac{(P^0)^3}{M}i\vec\sigma
\left([\vec E_T\times[\vec q'\times\vec B']]
-\frac{\omega^2}{\vec q^2}[\vec E'\times[\vec q\times\vec B]]\right)\nn\\
&&\rho_5=P^0i(\vec\sigma\vec B')(\vec q\vec E)\nn\\
&&\;\;+ P^0i\vec\sigma\cdot
\left([\vec E'\times[\vec q\times\vec B]]
+[\vec E_T\times[\vec q'\times\vec B']]\right)\nn\\
&&\rho_6=8MP^0
\left[i(\vec\sigma\vec B')(\vec q'\vec E)
-i(\vec\sigma\vec B)(\vec q\vec E')\right]\nn\\
&&\;\;+4P^0\vec q^2(\vec E_L\vec E')\nn\\
&&\nu\rho_{10}=4M\nu i\vec\sigma\cdot
\left([\vec E'\times\vec E]-[\vec B'\times\vec B]\right)\nn\\
&&\rho_{11}-4\rho_{5}=
-4Mi(\vec\sigma\vec\Delta)(\vec E\vec B'+\vec E'\vec B)\nn\\
&&\rho_{12}=4MP^0i(\vec\sigma\vec B')(\vec q\vec E)
-2P^0\vec q^2(\vec E_L\vec E')\nn\\
&&\;\;+\frac{Q^2}{8M}(\rho_{11}-4\rho_{5})\nn
\eeqn
where the higher order terms in $\omega$ were omitted, and
the use was made of the relations $\omega\vec B=[\vec q\times\vec E_T]$, 
$\vec E_T=-\frac{\omega}{\vec q^2}[\vec q\times\vec B]$, and 
$\vec E_L=(\hat q\vec E)\hat q$, with the unit vector along the direction of 
the virtual photon $\hat q$. \footnote{To simplify the above expressions, 
recoil corrections were neglected in tensors $f_{5,12}$ as i.e. 
$1-\frac{\Delta^2}{8MP^2}\approx1$, but not as 
$0-\frac{\Delta^2}{8MP^2}\approx0$} Appendix A enlists relations with the 
nucleon spinors that were used to derive the above results.

First, consider the similar structures 
$i\vec\sigma\cdot[\vec E'\times[\vec q\times\vec B]]$, 
$i\vec\sigma\cdot[\vec E_T\times[\vec q'\times\vec B']$
that enter $\nu\rho_4,\rho_5$. For both photons real, they reduce to 
$\omega i\vec\sigma\cdot[\vec E'\times\vec E]$ and come with the opposite sign, 
so they exactly cancel in $\rho_5$, and double in $\rho_4$. For VCS, it is 
only the first of the two that is leading order, thus $\rho_5$ obtains a 
contribution at leading order but in $\rho_4$ it is multiplied by $\omega^2$, 
and this amplitude is subleading in VCS. The possible way out is to build a 
third structure with the needed limit for RCS and for VCS, that would thus 
interpolate between the two low energy limits. This structure is 
\beqn
o_1&=&i\vec\sigma\cdot
\left([\vec E\times[\vec q'\times\vec B']]
-[\vec E'\times[\vec q\times\vec B]]\right)
\eeqn
\indent
The other two structures that do not match in 
low energy RCS and VCS are $\omega i\vec\sigma\cdot[\vec B'\times\vec B]$ and
$i(\vec\sigma\vec B')(\vec q\vec E)$, the first being part of $\rho_{10}$, 
and the second of $\rho_{12}$. 
The first tensor is purely transverse, being magnetic, whereas the second one 
is purely longitudinal with respect to the virtual photon. Once again, I will 
be looking for an interpolating tensor. The tensor of interest is
\beqn
o_2&=&i\vec\sigma\cdot
\left([\vec q\times[\vec E\times\vec B']]
-[\vec q'\times[\vec E'\times\vec B]]\right)
\eeqn
\indent
If the structures $o_{1,2}$ can be represented in a covariant form without 
introducing 
any spurious singularity, the low energy limit of the VCS amplitude 
will be related to the new, universal set of polarizabilities that are defined 
for real 
and virtual photons. This amounts in building four linear combinations out of 
$\rho_{4,5,10,12}$ such that two of them give the right limit at low energies, 
whereas the other two should be subleading in LEX. Such combinations are
\beqn
\phi_1&=&2\rho_5\,-\,\nu\rho_{10}\nn\\
\phi_2&=&2\nu\rho_5\,+\,Q^2\rho_{10}\nn\\
\phi_3&=&4M\rho_5-\tilde\rho_{12}-\nu\rho_{4}\nn\\
\phi_4&=&-\nu\tilde\rho_{12}+Q^2\rho_{4},
\eeqn
where the combination $\tilde\rho_{12}=\rho_{12}+\frac{Q^2}{4P^2}\rho_2-\frac{Q^2}{8M}(\rho_{11}-4\rho_5)$ was introduced in order to localize the structure of 
interest, $i(\vec\sigma\vec B')(\vec q\vec E)$.

\section{VCS observables}
To recollect, the leading contribution of the GP's to the VCS cross section 
arises as an interference between the divergent $\sim1/\omega$ parts of the 
Bethe-Heitler and FVCS Born amplitudes, and the first, $\sim\omega$ term in 
energy expansion of FVCS non-Born amplitude,
\beqn
a^{GP}_0&=&\sum_{spins}
\left[(T^{-1}_{BH}+T_{FVCS}^{B,-1})^*T_{FVCS,1}^{NB}\right.\nn\\
&+&\left.(T^{-1}_{BH}+T_{FVCS}^{B,-1})T_{FVCS,1}^{NB*}\right]
\label{b1}
\eeqn
\indent
The divergent parts are given by
\beqn
T_{BH}^{-1}&=&-\frac{e^3}{t}\bar u(p')\Gamma_\mu(\Delta)u(p)
\bar u(k')\gamma^\mu u(k)\nn\\
&\times&\left[\frac{{k'}^\nu}{(k'q')}-\frac{{k}^\nu}{(kq')}\right]
{\varepsilon'}^*_\nu\nn\\
T_{FVCS}^{B,-1}&=&-\frac{e^3}{Q^2}\bar u(p')\Gamma_\mu(q)u(p)
\bar u(k')\gamma^\mu u(k)\nn\\
&\times&\left[\frac{{p'}^\nu}{(p'q')}-\frac{{p}^\nu}{(pq')}\right]
{\varepsilon'}^*_\nu,
\eeqn
and the leading term of the non-Born part is given by 
\beqn
T_{FVCS}^{NB,1}=\frac{e}{Q^2}
\bar u(k')\gamma_\mu u(k)
\bar u(p')\sum_i\rho_i^{\mu\nu}f_i(Q^2)u(p){\varepsilon'}^*_\nu\nn\\
\eeqn
with $f_i(Q^2)$ related to the GP's as in Eq.(\ref{eq:GPs}). 
Note that for the leading terms $t$ can be substituted by $-Q^2$ since 
$t+Q^2=-2(qq')\sim\omega$. Unlike in \cite{guichon}, I perform the sum 
over spins in Eq. (\ref{b1}) in the covariant form. 
For the unpolarized case that is studied here, the calculation involves the 
trace 
\beqn
&&\frac{1}{4}\sum_{spins}\left[(T_{BH}+T_{FVCS}^B)_{-1}^*T_{FVCS,1}^{NB}
\right.\nn\\
&&\left.+(T_{BH}+T_{FVCS}^B)_{-1}T_{FVCS,1}^{NB*}\right]\\
&&=\frac{e^4}{(Q^2)^2}l_{\mu\mu'}H^{\mu\mu'\nu}
\left[\frac{{k'}_\nu}{(k'q')}-\frac{{k}_\nu}{(kq')}
-\frac{{p'}_\nu}{(p'q')}+\frac{{p}_\nu}{(pq')}\right]\nn,
\eeqn
with the usual unpolarised lepton tensor 
\beqn
l_{\mu\mu'}&=&\frac{1}{2}Tr[\sk'\gamma_{\mu'}\sk\gamma_\mu]\nn\\
&=&2(k_\mu k'_{\mu'}+k'_\mu k_{\mu'}-(kk')g_{\mu\mu'}),
\eeqn 
and the hadronic tensor is given by 
\beqn
H^{\mu\mu'\nu}
=\frac{1}{2}Tr[(\sp'+M)\Gamma^{\mu'}(q)(\sp+M)\sum_if_i\rho_i^{\mu\nu}]. 
\eeqn
\indent
There are only three distinct Dirac structures in the VCS tensor of Eqs. 
(\ref{eq:tensor1},\ref{eq:tensor2}), namely 
$1,\gamma^\beta\gamma_5$, and $\sigma^{\alpha\beta}$.
I rewrite the VCS tensor as 
\beqn
\sum_if_i\rho_i^{\mu\nu}=A^{\mu\nu}+B^{\mu\nu\beta}i\gamma_5\gamma_\beta
+C^{\mu\nu\alpha\beta}i\sigma_{\alpha\beta}
\eeqn
and the coefficients at these Dirac structures can be found analyzing 
Eq. (\ref{eq:EFT}), 
\begin{widetext}
\beqn
A^{\mu\nu}&=&f_1(-(qq')g^{\mu\nu}+{q'}^\mu q^\nu)
+4f_2P^{\mu}((Pq')q^\nu-(qq')P^{\nu})\nn\\
B^{\mu\nu\beta}&=&\frac{1}{2}\epsilon^{\lambda\sigma\nu\beta}q'_\sigma
(q^\mu q_\lambda-q^2g^\mu_\lambda)\hat f_5
+\frac{1}{2}\epsilon^{\lambda\sigma\mu\beta}q_\sigma
((qq')g^\nu_\lambda-q^\nu q'_\lambda)\hat f_{11}\nn\\
C^{\mu\nu\alpha\beta}&=&4q^\alpha g^{\mu\beta}
((Pq')q^\nu-(qq')P^{\nu})\hat f_6
-2{q'}^\alpha g^{\nu\beta}((Pq)q^\mu-q^2P^{\mu})f_{12}
\eeqn
\end{widetext}
where the shorthands were introduced, $\hat f_5=f_5-\frac{Q^2}{2M}f_{12}$, 
$2\hat f_6=2f_6+Q^2f_9$, and $\hat f_{11}=Q^2f_7+4f_{11}+\frac{Q^2}{2M}f_{12}$.
The three traces that  have to be computed, are
\begin{widetext}
\beqn
\frac{1}{2}Tr[(\sp'+M)\Gamma^{\mu'}(q)(\sp+M)]&=&4MP^{\mu'}G_E(Q^2)\nn\\
\frac{1}{2}Tr[(\sp'+M)\Gamma^{\mu'}(q)(\sp+M)i\gamma_5\gamma^\beta]&=&
-2G_M(Q^2)\epsilon^{\sigma\mu'\lambda\beta}P_\sigma\Delta_\lambda\nn\\
\frac{1}{2}Tr[(\sp'+M)\Gamma^{\mu'}(q)(\sp+M)i\sigma^{\alpha\beta}]&=&
\frac{2}{M}F_2(Q^2)P^{\mu'}[P^\alpha\Delta^\beta-P^\beta\Delta^\alpha]
+2MG_M(Q^2)[\Delta^\alpha g^{\mu'\beta}-\Delta^\beta g^{\mu'\alpha}]
\eeqn
\end{widetext}
\indent
Performing now Lorentz contraction, the hadronic tensor takes the following 
form
\begin{widetext}
\beqn
H^{\mu'\mu\nu}&=&((Pq')q^\nu-(qq')P^{\nu})
\left\{4MP^\mu P^{\mu'}\left[4G_Ef_2+\frac{Q^2}{M^2}F_2(2\hat f_6-f_{12})\right]
-Q^2g^{\mu\mu'}G_M(\hat f_5+\hat f_{11}+8M\hat f_6)\right\}\nn\\
&+&P^{\mu'}(-(qq')g^{\mu\nu}+{q'}^\mu q^\nu)
\left[4MG_Ef_1+Q^2G_M(\hat f_5+\hat f_{11}+4Mf_{12})\right]
\eeqn
\end{widetext}
To proceed, the electron kinematics needs to be defined. It is, in the case of 
the massless electron,
\beqn
k^\mu&=&E_e(1,\sin\theta'\cos\phi,\sin\theta'\sin\phi,\cos\theta'),\nn\\
k'^\mu&=&k^\mu-q^\mu.
\eeqn
with $E_e$ the energy of the initial electron, and $E_e'=E_e-\omega$ that of 
the final one. 
For the purpose of calculating the contribution of the GP's to the VCS 
observable to leading order, it is enough to take the electron kinematics in 
the limit of zero energy of the final photon, that is the kinematics of the 
elastic electron-proton scattering, $q\approx\Delta$. Then, one finds
\beqn
E_e&=&E_e',\nn\\
\cos\theta'&=&\sqrt{\frac{1-\epsilon}{1+\epsilon}}\nn\\
\sin\theta'&=&\sqrt{\frac{2\epsilon}{1+\epsilon}}
\eeqn
\indent
Above, I introduced the photon's longitudinal polarization parameter $\epsilon$,
\beqn
\epsilon&=&\frac{(E_e+E_e')^2-(\vec k-\vec k')^2}{(E_e+E_e')^2+(\vec k-\vec k')^2}.
\eeqn
\indent
The kinematical factors introduced in Section \ref{sec:obs} can be cast in the 
following form, after a straightforward but somewhat lengthy algebra:
\begin{widetext}
\beqn
v_1&\equiv&\frac{1}{Q^2}(Pq' q^\nu-qq' P^\nu)
\left[\frac{{k'}_\nu}{(k'q')}-\frac{{k}_\nu}{(kq')}
-\frac{{p'}_\nu}{(p'q')}+\frac{{p}_\nu}{(pq')}\right]\nn\\
&=&\sin\theta\left[\frac{4P^2\sin\theta}{4P^2-Q^2\cos^2\theta}
+\frac{P^0}{E_e}\frac{\sin\theta-\sqrt{\frac{2\epsilon}{1+\epsilon}}\cos\phi}
{(\sin\theta-\sqrt{\frac{2\epsilon}{1+\epsilon}}\cos\phi)^2+\cos^2\theta\sin^2\phi}\right]\nn\\
v_2&\equiv&\frac{1-\epsilon}{2Q^4}({q'}^\mu q^\nu-qq' g^{\mu\nu})P^{\mu'}
l_{\mu\mu'}
\left[\frac{{k'}_\nu}{(k'q')}-\frac{{k}_\nu}{(kq')}
-\frac{{p'}_\nu}{(p'q')}+\frac{{p}_\nu}{(pq')}\right]-\epsilon v_1\nn\\
&=&-\sqrt{\epsilon\frac{1+\epsilon}{2}}
\left[\frac{\cos\phi}{\sin\theta}v_1-\frac{P^0}{E_e}\frac{2\epsilon}{1+\epsilon}
\frac{\cos^2\theta\sin^2\phi}{(\sin\theta-\sqrt{\frac{2\epsilon}{1+\epsilon}}\cos\phi)^2+\cos^2\theta\sin^2\phi}\right].
\eeqn
\end{widetext}

\end{appendix}

\end{document}